\documentclass[lettersize,journal]{IEEEtran}
\usepackage{amsmath,amsfonts}
\usepackage[linesnumbered,ruled,vlined]{algorithm2e}
\usepackage{array}
\usepackage{textcomp}
\usepackage{stfloats}
\usepackage{url}
\usepackage{verbatim}
\usepackage{graphicx}
\usepackage{cite}
\usepackage{amsmath,amsfonts}
\usepackage{graphicx}
\usepackage{hyperref}
\usepackage{textcomp}
\usepackage{xcolor}
\usepackage{amssymb}
\usepackage{booktabs}
\usepackage{makecell}
\usepackage{listings}

\usepackage{algpseudocode}

\usepackage{multirow}
\usepackage{caption}
\usepackage{subcaption}
\usepackage{framed}
\usepackage{soul}
\usepackage[utf8]{inputenc}
\usepackage[T1]{fontenc}
\usepackage{microtype}
\usepackage{rotating}
\usepackage{paralist}
\usepackage{tabularx}
\usepackage{multicol}
\usepackage{pbox}
\usepackage{enumitem}	
\usepackage{colortbl}
\usepackage{pifont}
\usepackage{xspace}
\usepackage{url}
\usepackage{tikz}
\usepackage{fontawesome}
\usepackage{lscape}
\usepackage{color}
\usepackage{anyfontsize}
\usepackage{comment}
\usepackage{soul}
\usepackage{gensymb}
\usepackage{multibib}
\usepackage{balance}
\usepackage{footmisc}
\usepackage{xspace}
\usepackage[most]{tcolorbox}

\algnewcommand\Input{\item[\textbf{Input:}]}%
\algnewcommand\Output{\item[\textbf{Output:}]}%

\def\BibTeX{{\rm B\kern-.05em{\sc i\kern-.025em b}\kern-.08em
    T\kern-.1667em\lower.7ex\hbox{E}\kern-.125emX}}

\newcommand{\dmm}{\textsc{DeepMutation}\xspace}
\newcommand{\dm}{\textsc{DeepMutation++}\xspace}
\newcommand{\dc}{\textsc{DeepCrime}\xspace}
\newcommand{\name}{\textsc{MuFF}\xspace}

\definecolor{darkblue}{rgb}{0.0,0.0,0.6}

\newcommand{\algorithmicbreak}{\textbf{break}}
\newcommand{\BREAK}{\algorithmicbreak}

\newboolean{showcomments}
\setboolean{showcomments}{true}
\ifthenelse{\boolean{showcomments}}
  {\newcommand{\nb}[2]{
  \fbox{\bfseries\sffamily\scriptsize#1}
     {\sf\small$\blacktriangleright$\textit{\textcolor{red}{#2}}$\blacktriangleleft$}
   }
  }
  {\newcommand{\nb}[2]{}
   
  }

\newcommand\new[1]{{\color{black}#1}}

\definecolor{codegreen}{RGB}{0, 160, 0}
\definecolor{codered}{RGB}{160, 0, 0}

\begin{document}

\title{\name: Stable and Sensitive Post-training Mutation Testing for Deep Learning}

\author{Jinhan Kim, Nargiz Humbatova, Gunel Jahangirova, Shin Yoo, and Paolo Tonella
\thanks{J. Kim, N. Humbatova, and P. Tonella are with Software Institute, Università della Svizzera italiana, Lugano, Switzerland}
\thanks{G. Jahangirova is with Department of Informatics, King's College London, London, UK}
\thanks{S. Yoo is with School of Computing, KAIST, Daejeon, Republic of Korea.}
}

\markboth{Journal of \LaTeX\ Class Files,~Vol.~14, No.~8, August~2021}%
{Shell \MakeLowercase{\textit{et al.}}: A Sample Article Using IEEEtran.cls for IEEE Journals}

\maketitle

\begin{abstract}
Rapid adoptions of Deep Learning (DL) in a broad range of fields led to the development of specialised testing techniques for DL systems, including DL mutation testing. However, existing post-training DL mutation techniques often generate unstable mutants across multiple training repetitions and multiple applications of the same mutation operator. Additionally, while extremely efficient, they generate mutants without taking into account the mutants' sensitivity and killability, resulting in a large number of ineffective mutants compared to pre-training mutants. 
In this paper, we present a new efficient post-training DL mutation technique, named \name, designed to ensure the stability of the mutants and capable of generating killable and sensitive mutants. 
\name implements an automated stability check and introduces two mutation operators, named weight and neuron inhibitors. Our extensive empirical experiments show that \name generates mutants with 60\%pt and 25\%pt higher sensitivity compared to \dm and \dc, respectively, while also producing mutants that are more stable than those of \dm and different from the mutants of \dc. Moreover, \name preserves the benefits of the post-training mutation technique, being 61 times faster than \dc in generating mutants.
\end{abstract}


\section{Introduction}
\label{sec:intro}

\IEEEPARstart{D}{eep} Learning (DL) has significantly transformed the landscape of machine learning and artificial intelligence, becoming prevalent in numerous application domains~\cite{chen2015deepdriving, Jean2015aa, cui2018detection}. Correspondingly, it has become fundamentally crucial to ascertain the reliability of systems adopting DL, especially in safety-critical environments~\cite{varoquaux2022machine, Goodfellow43405}. This need has given rise to a number of testing techniques specifically designed for DL systems, including coverage criteria~\cite{Pei2017qy, Ma2018aa}, test adequacy metrics~\cite{Kim2019aa, kim2023evaluating}, and test generators~\cite{riccio2021deepmetis, tahereh2023deepatash}. 

The concept of mutation testing has also been successfully adapted in DL testing, leading to a variety of mutation techniques for DL systems~\cite{NHGJPT21, munn, deepmut, deepmut++}. One category is \textit{post-training} mutation techniques such as \dm~\cite{deepmut++}, which modify trained model's weights or structure using Mutation Operators (MOs) like weight shuffling or layer duplication. These techniques have been regarded as efficient tools for generating DL mutants~\cite{deepmut}, although they inject faults that do not mimic real faults~\cite{NHGJPT21}.

In contrast, \textit{pre-training} mutation techniques simulate the occurrence of real faults by mutating the source code or the training data of the model before the training process. \dc~\cite{NHGJPT21} implements this approach, proposing MOs like deleting a portion of training data or modifying the number of training epochs. While \dc's MOs are more realistic as they were extracted from real faults~\cite{taxonomy}, \dc also suffers from high costs associated with retraining a model from scratch for the generation of a mutant.

For both pre- and post-training mutation techniques, when applying them to DL models, an important factor to consider is the inherent stochasticity of the resulting mutants. The MO applied to the same model may result in different mutants, leading to significantly different mutation scores. This stochasticity arises from two sources: randomness in the training process and inherent randomness of MOs employed by mutation tools (e.g., randomly selecting layers or training data elements to mutate). To address this issue, \dc leveraged the concept of statistical killing proposed by Jahangirova \& Tonella~\cite{jahangirovantonella}. This concept is based on training both the mutated and the original (non-mutated)  models multiple times and checking whether the difference in their performance (e.g., accuracy) is statistically significant. In turn, this procedure requires a sample of performance values that is sufficient to ensure a stable performance estimation across repeated experiments, as unstable performance estimates may lead to inconsistent and unreliable mutation scores.  However, no existing post-training mutation tool has considered the stochasticity of the training process and the MOs, as well as the stability of the corresponding performance estimates, when calculating the mutation score.

\begin{figure}[!t]
  \centering
  \begin{subfigure}{0.48\linewidth}
      \centering
      \includegraphics[width=0.77\linewidth]{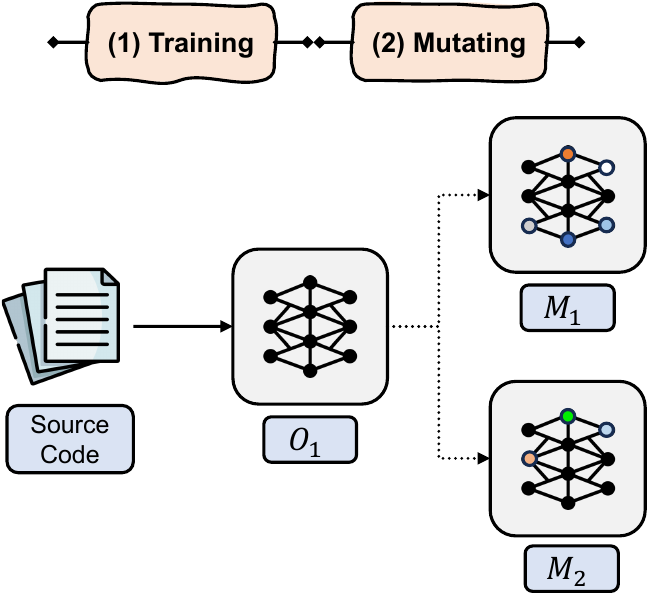}
      \caption{Case 1}%
      \label{fig:ex1}
  \end{subfigure}%
  \begin{subfigure}{0.48\linewidth}
      \centering
      \includegraphics[width=0.77\linewidth]{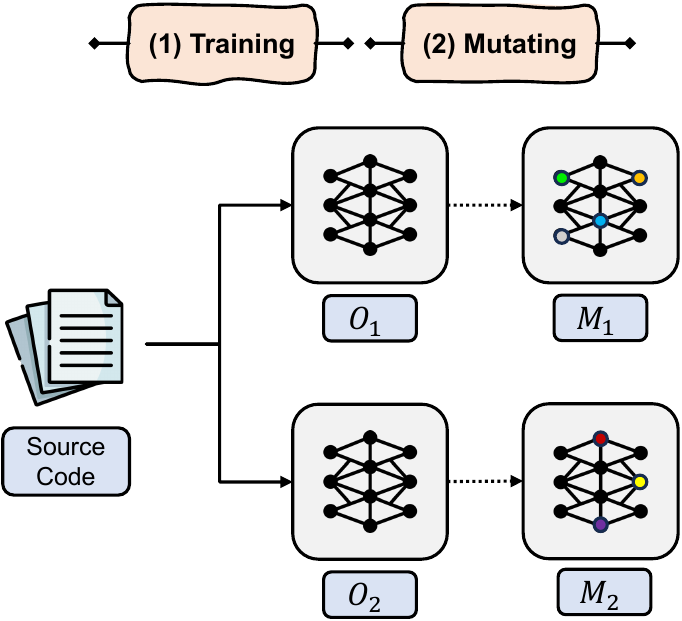}
      \caption{Case 2}%
      \label{fig:ex2}
  \end{subfigure}
  \caption{This example depicts one (a) or two (b) original DL models, $O_1$ and $O_2$, trained individually using the same source code and hyperparameters. Then, two mutants, $M_1$ and $M_2$, are generated by applying the same post-training mutation operator either to $O_1$ only, or to both $O_1$ and $O_2$.}
  \label{fig:motivating_example}
\end{figure}

To illustrate this issue in the existing post-training mutation tool \dm, let us consider the example shown in Figure~\ref{fig:motivating_example}. $O_1$ and $O_2$ are two models trained on the same MNIST dataset with the same source code (i.e., the same model structure and hyperparameters). From these, we can generate two mutants, $M_1$ and $M_2$, in two ways: by applying \dm's Neuron Activation Inverse operator to $O_1$ two times (Figure~\ref{fig:ex1}), or by individually applying the same MO to each of $O_1$ and $O_2$ (Figure~\ref{fig:ex2}). In both cases, we observe a significant difference between the accuracies of $M_1$ and $M_2$. In the first case, the difference is 44.23\%pt while in the second case, it is 62.86\%pt.
\new{Moreover, the rate of disagreements\footnote{The proportion of training inputs where the mutant instances yield inconsistent outcomes.} varies significantly between the two cases: in the first case, the disagreement rate is 32.21\%, whereas it is 65.89\% in the second. This exemplifies the inherent stochasticity present in post-training mutants and the risk of instability when estimating a model's performance and mutation score.}

\new{This paper first rigorously investigates the stochastic nature of post-training mutations with two initial research questions (RQs 1 \& 2) in Section~\ref{sec:study1}. 
The empirical analysis of the stochastic outputs of existing post-training mutation (Section~\ref{sec:study1}) consists of two research questions:
RQ1 evaluates the disagreement rate of mutants across four subject models and datasets, while RQ2 assesses the stability of their performance metrics, such as accuracy, as the number of mutant instances increases.\footnote{Throughout this paper, the term \textit{instance} denotes a model with the specific weights obtained after a given training session or after applying a post-training mutation.} The goal of these initial sections is to empirically validate our intuition on the need for a probabilistic notion of mutation killing and to provide insights to guide the design of our new mutation tool.}

Subsequently, based on the empirical findings that highlight the instability and low sensitivity of existing post-training mutation tools \new{(Section~\ref{sec:study1})}, \new{Section~\ref{sec:new_tool} proposes \name (DL \underline{Mu}tation, \underline{F}ast \& \underline{F}urious)}, a novel \textit{post-training} mutation technique. To address the instability of post-training mutants, \name employs an automatic stability-checking procedure. This procedure iteratively applies MO to multiple original models, generating several instances until the estimation of performance (e.g., accuracy) converges to a stable value with a low standard error. Additionally, \name utilises a binary search to find MO configurations that maximise the generation of killable mutants, while avoiding unstable or trivial mutants. 

\new{Previous empirical studies have shown that \dm mutants demonstrate lower sensitivity to the quality of the test set when compared to \dc's pre-training mutants~\cite{NHGJPT21}. \textit{Sensitivity}, a metric previously adopted as a proxy for mutation quality~\cite{NHGJPT21}, measures a mutant's responsiveness to the quality of the test set. It assesses how much the mutant's behaviour changes when evaluated with stronger or weaker tests, calculated as the relative difference in mutation scores between strong and weak test sets, normalised by the strong set's score. High sensitivity indicates a mutant's behaviour is greatly influenced by test effectiveness, hence helping developers identify areas for test improvement. Through RQ3 (Section~\ref{sec:new_tool}), we explore the low sensitivity of \dm mutants and investigate the underlying causes.}

To address the limited sensitivity of existing post-training mutants, \name introduces two new mutation operators called Weight Inhibitor (WI) and Neuron Inhibitor (NI). They reduce the magnitude of weight values, with fine and smooth control on the reduction of the signal transmitted by the mutated weight between neurons. This nuanced approach promotes both stability and efficacy in DL mutation. 
Combined with the binary search of killable MO configurations, the new operator ensures that killable but challenging -- hence, highly sensitive -- mutants are generated.

We conduct an extensive empirical study \new{(Section~\ref{sec:study2})} comparing \name to the state-of-the-art tools, \dm and \dc, across four models and datasets. Our results show that \name's mutants are significantly more sensitive to the test set quality, with on average 60\%pt and 25\%pt increase over \dm and \dc, respectively. 
At the same time, \name achieves a substantial speedup over \dc, generating mutants 61 times faster.
Finally, the results of spectral analysis indicate that \name's mutants hold unique characteristics, distinguishing them from \dc's mutants.

\section{Background}
\label{sec:background}

DL mutation tools can generally be classified into two main categories: post-training mutations and pre-training mutations. In the following, we introduce representative tools of each type and highlight their varying assumptions and definitions.

\begin{table}[htb]
\centering
\caption{Mutation Operators of \dm}\label{tab:dm_operators}

\scalebox{0.93}{
\begin{scriptsize}
\begin{tabular}{ll}
\toprule
Operator (ID) & Description \\
\midrule
Gaussian Fuzzing (GF) & Introduces Gaussian noise by multiplying \\ & the noise with the weights. \\
Weight Shuffle (WS) & Randomly shuffles the weights of a neuron. \\
Neuron Effect Block (NEB) & Blocks a neuron's influence on its connected \\ & neurons. \\
Neuron Activation Inverse (NAI) & Inverts the activation of the neuron. \\
Neuron Switch (NS) & Exchanges two randomly selected neurons \\ & within the same layer. \\
Layer Removal (LR) & Removes a randomly chosen layer from the model. \\
Layer Addition (LA) & Inserts a random activation layer into the model. \\
Layer Duplication (LD) & Creates a duplicate of a randomly selected layer. \\

\bottomrule
\end{tabular}
\end{scriptsize}
}
\end{table}

\subsection{Post-training Mutations}
\label{sec:post-training}
Post-training mutations modify the weights or structure of a trained model and such changes typically result in a degradation of the model's performance\footnote{However, on occasion, these changes can improve or \textit{fix} the model, as proved in the DL repair techniques~\cite{Sohn2022cr, li2023adaptive}. This duality between automated program repair and mutation testing has been highlighted in the literature~\cite{Weimer2013ma, Kim2023pa}.}.
\dm~\cite{deepmut++} is a state-of-the-art post-training mutation tool that directly mutates the trained weights, neurons, or layers using mutation operators like Neuron Activation Inverse (NAI) that inverts the activation status of a neuron. A full list of its operators is shown in Table~\ref{tab:dm_operators}.
The advantage of performing direct mutations on the weights is high efficiency, as it circumvents the need for retraining the model and allows for a quick generation of an extensive set of mutants. 
Nevertheless, since their weight- or neuron-level operators modify the weights in a manner that is not interpretable by humans, it remains a challenge to identify any connection to real DL faults~\cite{taxonomy}.
Moreover, \dm lacks any guidance for  mutant generation to encourage the creation of killable and non-trivial mutants,
where a mutant is intuitively deemed \textit{killable} if there exists a test set that causes a significant accuracy drop w.r.t. the original model, while it is \textit{trivial} if almost any test set can kill it (more precise definitions follow below).
For example, drastic changes such as inversion of the activation of the neurons can result in the generation of mutants with a very low accuracy. These mutants are likely to be killable, but at the same time they are also likely trivial. While \dm has a user-specified threshold to discard possibly trivial mutants based on their accuracy, this parameter is not easy to set manually, as it may either cause \dm to fail to generate any mutants (i.e., all generated mutants fall below the threshold), or to generate only \textit{equivalent} mutants, i.e., non-killable mutants that are indistinguishable from the original model.

\dm defines its mutation score based on class-level prediction errors: A mutant $m_i$ is considered killed for class $c$ if the original model correctly predicts $c$ for any test case $t$, but $m_i$ does not. The score is calculated as:
\begin{equation}
    MS = \frac{\sum_{m_i \in M} |killedClasses(T, m_i)|}{|M| \times |C|}
\end{equation}
where $C$ is a set of all classes, $T$ is a test set, $M$ is a set of mutants, $killedClasses(T, m_i)$ counts the classes killed for $m_i$ by $T$. 
As pointed out in the previous section, this definition does not take into account the various sources of randomness that affect the creation of a mutant $m_i$, and moreover, it is only applicable to classification problems. 

\begin{table}[htb]
\centering
\caption{Mutation Operators of \dc}\label{tab:dc_operators}

\scalebox{0.95}{
\begin{scriptsize}
\begin{tabular}{ll|ll}
\toprule
ID & Description & ID & Description \\
\midrule
AAL & Add activation function to layer & RCP & Change patience parameter \\
ACH & Change activation function & RCW & Change weights regularisation \\
ARM & Remove activation function & RRW & Remove weights regularisation \\
HBS & Change batch size & TAN & Add noise to training data \\
HDB & Disable data batching & TCL & Change labels of training data \\
HLR & Decrease learning rate & TCO & Make output classes overlap \\
HNE & Change number of epochs & TRD & Remove portion of training data \\
LCH & Change loss function & TUD & Unbalance training data \\
OCG & Change gradient clipping & VRM & Remove validation set \\
OCH & Change optimisation function & WAB & Add bias to a layer \\
RAW & Add weights regularisation & WCI & Change weights initialisation \\
RCD & Change dropout rate & WRB & Remove bias from a layer \\
\bottomrule
\end{tabular}
\end{scriptsize}
}
\end{table}

\subsection{Pre-training Mutations}
\label{sec:pre-training}

Pre-training mutations share the core principle of traditional mutation testing by directly modifying the sources of a DL program. Unlike post-training mutations, however, training is needed to generate final mutants, which hinders its practical applicability as training usually requires considerable time and resources.
\dc~\cite{NHGJPT21} is one such pre-training mutation tool that aims to simulate real DL faults to generate realistic mutants~\cite{taxonomy}. Table~\ref{tab:dc_operators} outlines all mutation operators of \dc. For example, these operators include changing the activation function and removing a portion of the training data.

\dc accounts for the randomness stemming from model training, as training the same model multiple times can yield significant variance. This differs from \dm, which assumes that a mutant comprises a single instance. In contrast, \dc posits that a single mutant consists of several \textit{instances} of the trained model, all originating from the same mutated source. \new{For example, when \dc applies the `Change number of epochs (HNE)' operator, it mutates the original source code and subsequently trains this mutated version 20 times, treating the resulting 20 instances as a single mutant. This enables a statistical definition of mutant killing by comparing the accuracies (or other evaluation metrics) of the mutant instances and the original model instances~\cite{jahangirovantonella}. Note that this computation assumes the original model also consists of multiple instances, each trained separately using the original source code. If a statistical difference is observed between two sets of accuracies, it indicates that the mutant is killed:}

\begin{equation} \label{eq:kill}
  isKilled = 
  \begin{cases}
      True,& \text{if } \text{\textit{effectSize}}(\mathcal{A}_{M}(T), \mathcal{A}_{O}(T)) \ge \beta  \\
            & \text{ and } \text{\textit{p\_value}}(\mathcal{A}_{M}(T), \mathcal{A}_{O}(T)) < \alpha \\
      False,& \text{otherwise.}
  \end{cases}
\end{equation}

\noindent where $\mathcal{A}_{M}(T)$ and $\mathcal{A}_{O}(T)$ denote the accuracies of the mutant and original model instances, respectively, for a given test set $T$. $\beta$ and $\alpha$ are predefined thresholds for effect size and p-value, respectively,  which determine the significance of the accuracy difference. Building on this definition of statistical mutant killing, \dc's authors define their own mutation score for each mutation operator. Given a test set $T_{test}$, train set $T_{train}$, and a mutation operator $MO$, this mutation score is calculated as follows:
\begin{equation} \label{eq:ms}
    MS(MO, T_{test}) = \frac{\left| K(MO, T_{test})\cap K(MO,T_{train})) \right|}{\left| K(MO, T_{train}) \right|}
\end{equation}
where $K(MO, T_{test})$ and $K(MO, T_{train})$ represent sets of configurations of $MO$ that are killed by $T_{test}$ and $T_{train}$, respectively, while an \textit{MO configuration} is a specific choice of the values for the MO's parameters (e.g., a specific percentage of training data to remove for TRD). Here, $T_{train}$ is used in the denominator as mutants are anticipated to be most sensitive to the training set. Therefore, any configuration not covered by the $T_{train}$ is considered likely non-killable (i.e., likely equivalent) and excluded from the computation of the mutation score.
Let us consider the TAN mutation operator (see Table~\ref{tab:dc_operators}). If $T_{train}$ kills all configurations with the noise applied to 30\% or more of the train data, while $T_{test}$ requires at least 50\% of the train data to be affected by the noise, then the mutation score is $\frac{(1.0 - 0.5)}{(1.0 - 0.3)} \approx 0.71$. As our goal is to design a new post-training mutation tool that generates stable mutants with statistical evidence, we adopt \dc's definitions of mutation killing and mutation score.

\new{The quality of the mutant can be approximated using the sensitivity of the mutant to the quality of the test sets. Following Humbatova et al.~\cite{NHGJPT21}, we utilise the strong and weak test sets to compute such sensitivity. The strong test set is the original test set provided with the subject and the weak test set is artificially constructed by selectively removing test inputs from the strong set, retaining only those inputs for which the model under test exhibits high confidence. In the case of a regression problem, we create weak test sets by excluding inputs with low mean loss or low standard deviation of loss across multiple instances of the original models. With these two sets, we calculate sensitivity as follows:}
\begin{equation}\label{eq:sens}
    Sensitivity = \frac{MS(MO, T_{strong}) - MS(MO, T_{weak})}{MS(MO, T_{strong})}
\end{equation}
\noindent \new{where $MS(MO, T_{strong})$ and $MS(MO, T_{weak})$ represent the mutation scores of the mutation operator $MO$ on the strong and weak test sets, respectively. Higher sensitivity, indicated by values closer to 1.0, signifies that the mutation operator has generated high-quality mutants. These mutants are resistant to being killed by the weak test set but are effectively killed by both the strong and train sets. Humbatova et al.~\cite{NHGJPT21} empirically demonstrated that \dm mutants exhibit lower sensitivity compared to \dc mutants. In our study, we aim to investigate the reasons behind the low sensitivity of post-training mutants and utilise this understanding to design a new post-training mutation tool. Subsequently, we will compare the sensitivity of the mutation tools in RQ3.}

\section{Variability and Instability of Post-training Mutations}
\label{sec:study1}

To gather quantitative, objective evidence, as well as a deeper understanding of the occurrence of variability and instability in post-training DL mutation tools like \dm, we designed and conducted the following empirical study.

\begin{table*}[h]
\caption{Datasets and models}\label{tab:dataset}
\centering
\resizebox{\textwidth}{!}{%
\begin{tabular}{llllll}
    \toprule
    Dataset & Id & Task & Description & DL Model & Performance \\
    \midrule
    MNIST & MN & C & \makecell{Handwritten digit images including 60,000 images for training\\ and 10,000 images for testing.} & \makecell{Eight-layer convolutional \\neural network.} & 99.15\% (Accuracy) \\
    \midrule
    \makecell{Kaggle Speaker\\Recognition\\Dataset} & SR & C & \makecell{Audio files used to recognise and classify unique speakers,\\ including 5,401 training inputs and 1,350 test inputs.} &  \makecell{Residual network with 46\\ convolutional layers, followed \\by two dense layers.} & 99.44\% (Accuracy) \\
    \midrule
    Unity Eyes & UE & R & \makecell{Synthesised eye region images, used to map eye images \\and 2D head angles to eye gaze angles, including 103,428 \\and 25,857 inputs for training and testing. } & \makecell{Two convolution layers\\ with max-pooling and a dense- \\layer with 500 neurons.} & 0.05 (MSE) \\
    \midrule
    Udacity & UD & R & \makecell{Self-driving car dataset for predicting steering angles, \\including 9,792 and 2,432 inputs for training and testing.}  & Dave-2 from Nvidia & 0.01 (MSE)\\
    \bottomrule
\end{tabular}
}
\end{table*}

\subsection{Research Questions}

\subsubsection{\textbf{RQ1 (Variability)}} \textit{How much variance is introduced in post-training mutants by both the mutation operators of \dm and the stochasticity of model training?}

Variability in \dm mutants is influenced by either the training process (hence, we gather multiple trained instances of the original model before applying the mutation) or by the randomness associated with \dm's MOs. This RQ aims to quantify this variability across two scenarios: In the first, 20 mutant instances are generated from a single original model; in the second scenario, we generate one mutant instance from each of 20 original model instances, resulting in a total of 20 mutant instances. To measure variability, we use \textit{disagreement rate}, calculated as the percentage of training inputs where the 20 mutant instances yield inconsistent prediction outcomes, meaning that some mutant instances make correct predictions while others are wrong. 
We do not consider the case where all predictions are unique but wrong as a disagreement, because they do not contribute to the variability of the accuracy measurements (i.e., they all reduce accuracy).

\subsubsection{\textbf{RQ2 (Stability)}} \textit{How many instances are necessary to generate stable post-training mutants of \dm?} %

While RQ1 explores the extent of variability of predictions for individual inputs, RQ2 shifts the focus to the stability of the overall performance (e.g., accuracy) measurement as the number of instances grows. 
To assess stability, we calculate the Relative Standard Error (RSE) affecting the estimated mean of the evaluation metric (e.g., accuracy) measured on the given test data. 
Let $\mathcal{O}$ be a set of $m$ original model instances, all trained using the same code/dataset. Let us consider a mutant as composed of a set of instances, denoted as $I$, generated by applying a post-training MO $n$ times to each of the $m$ original models in $\mathcal{O}$, i.e., $|I| = n \times m$. Each instance $i \in I$ is evaluated under a given test set, yielding a set of $n \times m$ mutant evaluation metric values, $L_{I}$, with mean $\mu$ and standard deviation $\sigma$. RSE is calculated as the ratio of the Standard Error (SE) to the mean:

\begin{equation}
     \begin{aligned}
    SE = \frac{\sigma}{\sqrt{nm}} \hspace{1cm}
    RSE = \frac{SE}{\mu}
    \end{aligned}
\end{equation}

RSE conforms to the law of large numbers: it decreases at increasing $n$ or $m$. Hence, RSE helps determine the minimum number of instances, required to achieve a specific level of stability of the mutant (e.g., RSE lower than 5\%).

For this RQ2, we choose $m = 20$ and $n \in \{1, 2, 3, 4, 5\}$: we start with 20 original model instances and generate one mutant instance for each, yielding a total of 20 instances ($|I|$). We then iteratively generate additional mutant instances, adding 20 instances to $I$ in each step, by mutating again each of the 20 original model instances, until the mutant achieves stability according to a pre-defined RSE threshold (in our experiments, 5\%) or reaches a maximum of 100 instances. 

Some of \dm's mutation operators (namely, GF, WS, NEB, NAI, NS; see Table~\ref{tab:dm_operators}) can be configured with a \textit{ratio} parameter that specifies the proportion of neurons or weights to be mutated.
We also investigate how different ratios affect the stability of \dm mutants, by analyzing four ratio values: 0.01, 0.2, 0.6, and 1.0. %
Note that, as LR, LA, and LD (see Table~\ref{tab:dm_operators}) have no ratio parameter, we present their results without any specific selection (the layer they are applied to is selected randomly).

\subsection{Subjects and Configurations}
\label{sec:study1_configurations}

We employ four different pairs of models and datasets, as shown in Table~\ref{tab:dataset}. The `Task' column represents the type of problem solved by the model: classification (`C') or regression (`R'). DL models are the same previously used for evaluation of \dm and \dc~\cite{NHGJPT21}. MNIST (MN) is a widely used dataset of handwritten digits, along with an 8-layer convolutional network. Speaker Recognition (SR) dataset from Kaggle is used to recognise and classify speakers from audio files of speech recordings with the help of a 46-layer convolutional neural network. UnityEyes (UE) dataset contains various synthetic eye region images and labels, with a convolutional network (LeNet-5 architecture) that maps eye image and head rotation angle to eye gaze angle. %
Udacity (UD) dataset is derived from car driving simulations used to predict the steering angle, for the lane keeping task, with the Dave-2 model from Nvidia~\cite{bojarski2016end}. To decide if a prediction can be deemed correct in regression tasks, we use the pre-defined thresholds set at 5 degrees difference for UE and 0.3 for UD~\cite{NHGJPT21}.

For the RQs 1 \& 2, we use the default configuration of \dm to generate mutants, and the stability threshold of RSE is set to 5\%. We exclude LR and LA for SR, and LR, LA, and LD for UE as \dm fails to run successfully when applying these MOs to these specific subjects. We report the results based on five runs of the experiments. All experiments are conducted on systems equipped with Ubuntu 18.04, Intel Xeon E5-2630 v4, Nvidia TITAN Xp GPUs, and 256GB of RAM.

\subsection{Results}

\subsubsection{RQ1 (Variability)}
\label{sec:RQ1}

\begin{table*}[h]
\centering
\caption{Disagreement rate of predictions of 20 \dm mutants generated from one original model}\label{tab:rq1_1}
\scalebox{1.0}{
\begin{tabular}{c|c|c|c|c|c|c|c|c}
\toprule

Subj. & \multicolumn{8}{c}{Mutation Operator of \dm}  \\ 
& GF & WS & NEB & NAI & NS & LR & LA & LD  \\ 
 
\midrule

MN & 0.28\% & 67.11\% & 33.07\% & 99.97\% & 79.33\% & 0.00\% & 99.88\% & 0.00\% \\
SR & 8.63\% & 81.21\% & 70.17\% & 100.00\% & 45.27\% & N/A & N/A & 5.67\% \\
UE & 70.52\% & 98.04\% & 91.90\% & 98.30\% & 94.91\% & N/A & N/A & N/A  \\
UD & 33.24\% & 74.62\% & 30.74\% & 46.27\% & 42.12\% & 27.28\% & 99.98\% & 27.57\% \\
\bottomrule

\end{tabular}
}
\end{table*}

The disagreement rates for the first scenario (i.e., mutants generated from a single original model) are presented in Table~\ref{tab:rq1_1}, where each column represents a MO. The results show that the disagreement rates vary drastically depending on the MO and the subject. For example, NAI consistently exhibits high disagreement rates across subjects, suggesting that it is more likely to considerably mutate the original model, potentially introducing higher variability in the mutants. On the other hand, LR and LD show a 0.00\% rate in some subjects. This can be explained by the fact that LR only targets the layers that have the same input-output shape, which makes it consider only dropout layers. The removal of these dropout layers does not affect the model predictions.

\begin{table*}[h]
\centering
\caption{Disagreement rate of predictions of 20 \dm mutants generated across 20 original models}\label{tab:rq1_2}
\scalebox{1.0}{
\begin{tabular}{c|c|c|c|c|c|c|c|c}
\toprule

Subj.  & \multicolumn{8}{c}{Mutation Operator of \dm}  \\ 
&  GF & WS & NEB & NAI & NS & LR & LA & LD  \\ 

\midrule
MN & \underline{1.11}\% & \underline{77.11}\% & \underline{12.46}\% & \underline{96.67}\% & \underline{61.86}\% & 0.99\% & \underline{99.98}\% & 0.99\% \\
SR & \underline{12.81}\% & \underline{76.47}\% & \underline{47.45}\% & \underline{88.43}\% & \underline{83.06}\% & N/A & N/A & 10.46\% \\
UE & \underline{80.72}\% & \underline{98.65}\% & \underline{96.33}\% & \underline{98.69}\% & \underline{98.52}\% & N/A & N/A & N/A  \\
UD & \underline{31.31}\% & \underline{50.71}\% & \underline{47.47}\% & \underline{62.36}\% & \underline{55.33}\% & \underline{26.70}\% & \underline{100.00}\% & 26.45\% \\

\bottomrule

\end{tabular}
}
\end{table*}

Table~\ref{tab:rq1_2} presents the results of the second scenario (i.e., one mutant for each of the 20 original models). The disagreement rates that are larger than those between pairs of original models are underlined. We can observe similar trends as in the first scenario, with an average disagreement rate of 57.15\%. Among MOs, GF exhibits the disagreement rates closest to the original models, indicating its lower variability compared to the others. %

\begin{tcolorbox}[boxrule=0pt,frame hidden,sharp corners,enhanced,borderline north={1pt}{0pt}{black},borderline south={1pt}{0pt}{black},boxsep=2pt,left=2pt,right=2pt,top=2.5pt,bottom=2pt]
\textbf{Answer to RQ1}: Our findings show that the disagreement rates among \dm mutants are significantly high, indicating a high degree of variability of the predictions, especially for some specific mutation operators/subjects. 
\end{tcolorbox}

\subsubsection{RQ2 (Stability)}
\label{sec:RQ2}

\begin{figure}[!ht]
  \centering

  \begin{subfigure}{0.5\linewidth}
      \centering
      \includegraphics[width=\linewidth]{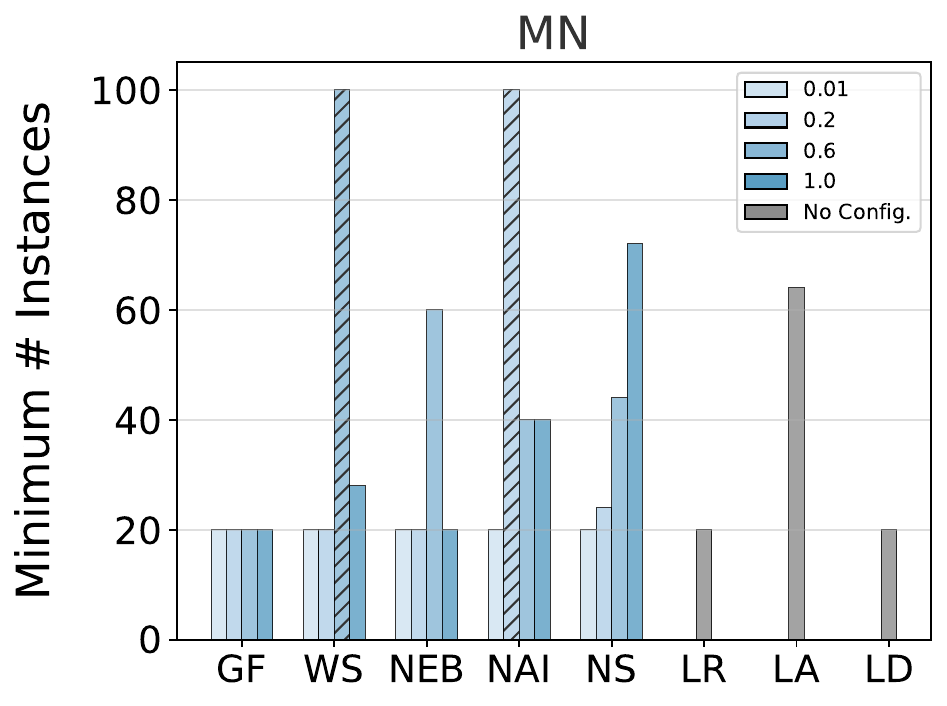}
      \label{fig:mnist_optimal}
  \end{subfigure}%
  \begin{subfigure}{0.5\linewidth}
      \centering
      \includegraphics[width=\linewidth]{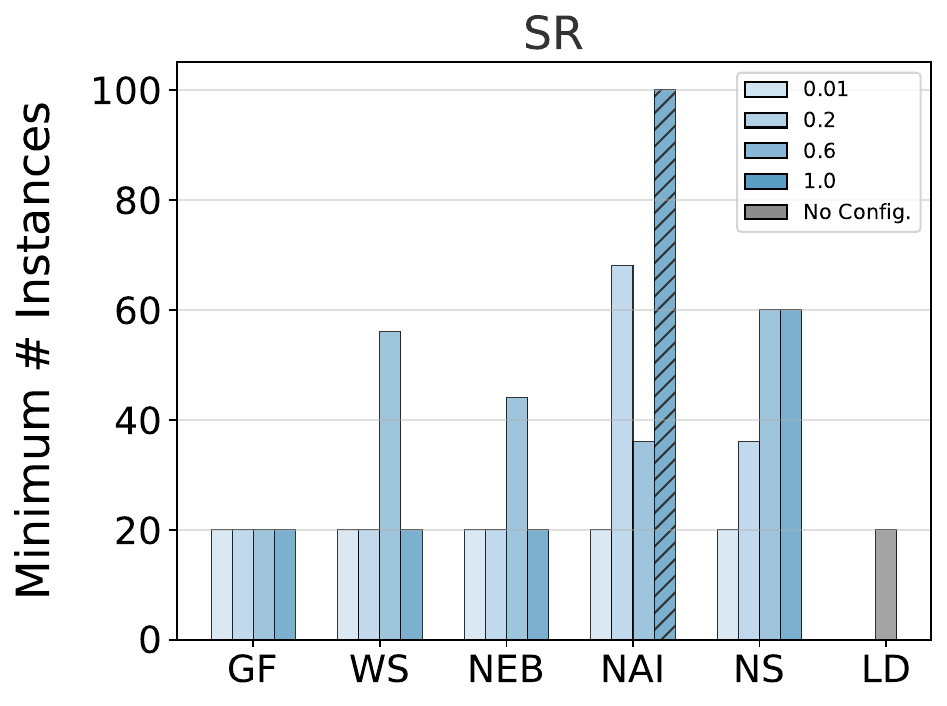}
      \label{fig:sr_optimal}
  \end{subfigure}
  \begin{subfigure}{0.5\linewidth}
      \centering
      \includegraphics[width=\linewidth]{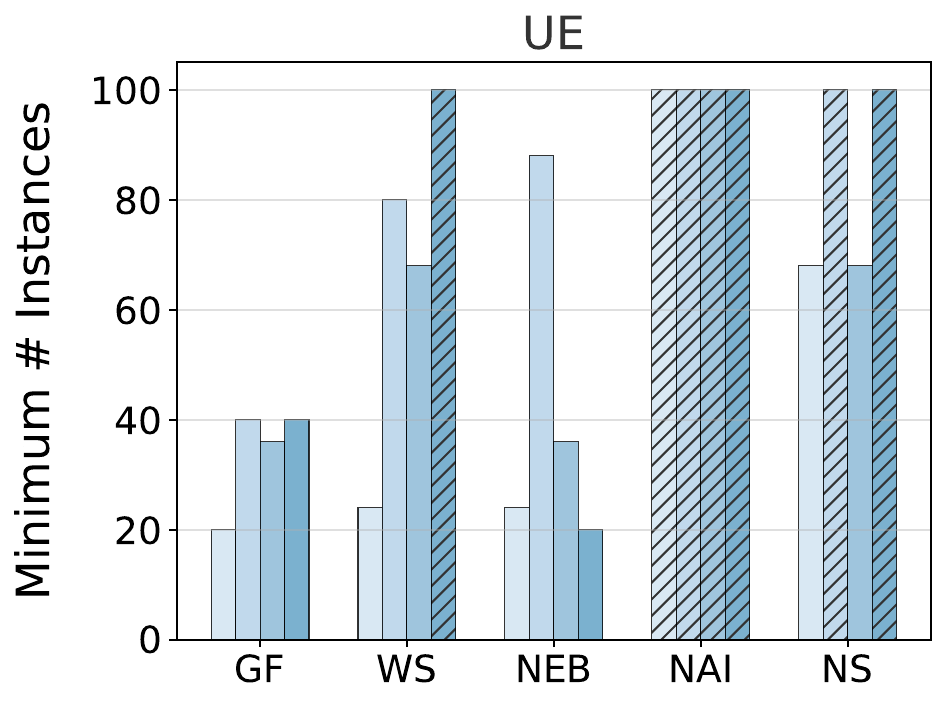}
      \label{fig:ue_optimal}
  \end{subfigure}%
  \begin{subfigure}{0.5\linewidth}
      \centering
      \includegraphics[width=\linewidth]{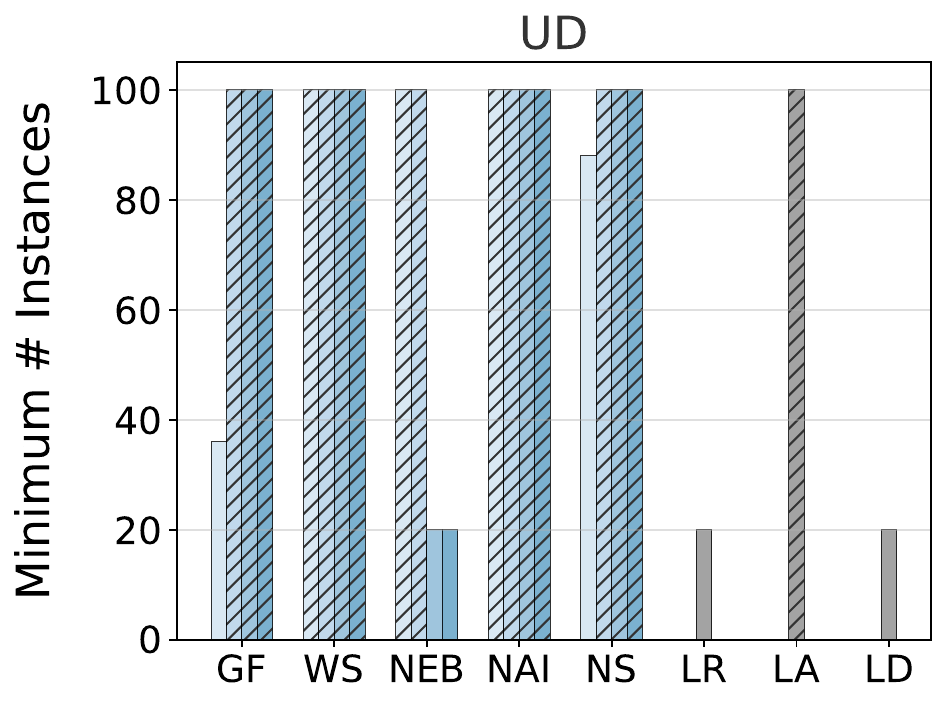}
      \label{fig:ud_optimal}
  \end{subfigure}
  \caption{Minimum number of instances for stable \dm mutants. Each bar colour corresponds to a specific MO configuration, i.e., ratio value.}
  \label{fig:optimal_num}
\end{figure}

The high variability of the predictions observed in RQ1 implies that, depending on the mutation operator and subject, the number of instances needed to reliably estimate the mutant's accuracy might be sometimes high.
Figure~\ref{fig:optimal_num} illustrates the minimum number of instances needed for generating stable \dm mutants. Bar colours represent different ratios, while striped bars indicate the cases in which we fail to find a stable mutant even after 100 instances. Overall, we observe significant variation in the number of instances required across subjects, MOs, and MO configurations.
The mutants on subjects MN and SR require relatively fewer instances than UE and UD. For the latter, most mutants remain unstable even after 100 instances. %
This is potentially due to the lower redundancy of their model structure, making them highly sensitive to weight mutations. %
We can observe that different configurations within the same MO could give vastly different outcomes, highlighting the need for an automated method that determines MO configuration and the number of instances necessary for generating stable mutants.

\begin{tcolorbox}[boxrule=0pt,frame hidden,sharp corners,enhanced,borderline north={1pt}{0pt}{black},borderline south={1pt}{0pt}{black},boxsep=2pt,left=2pt,right=2pt,top=2.5pt,bottom=2pt]
\textbf{Answer to RQ2}: The minimum number of instances for generating stable mutants varies significantly depending on the subject, mutation operator, and its configuration. This highlights the necessity of automating the selection of mutant configuration and number of instances. %
\end{tcolorbox}

\section{\name}
\label{sec:new_tool}

Our initial empirical study highlighted the variability and instability present in the state-of-the-art post-training mutation tool \dm. This section presents \name, an efficient post-training DL mutation technique that is stable and capable of producing mutants that are both killable and non-trivial. For stability, we take into account the various sources of randomness by addressing the stability of a mutant explicitly, in an automated manner. Specifically, we propose a new approach that determines the minimum number of instances required to yield stable mutants. To produce killable and non-trivial mutants, we adopt binary search to tune the values of MO parameters (e.g., the ratio of neurons affected by the MO), instead of manual choice used by existing tools. The binary search, combined with the stability measurement, allows us to accurately and reliably estimate both mutation score and mutation sensitivity, leading to the configurations of the given MO that can produce killable mutants.

To address the issue of low sensitivity of existing post-training mutation techniques~\cite{NHGJPT21} (i.e., they give the same or similar mutation scores when either a weak or a strong test set is used), \name introduces new post-training mutation operators, named Weight Inhibitor (WI) and Neuron Inhibitor (NI). They offer fine-grained control over the model weight changes, guiding the generation of challenging mutants, by means of binary search, which turn out to be highly sensitive. They are also stable, thanks to our automated stability checks.

\begin{algorithm}[ht]
  \caption{Binary Search}\label{algo:bs}
  \small
  \SetCommentSty{mycommfont}
  \SetKwInput{KwPrecondition}{Precondition}
  \SetKwInput{KwPostcondition}{Postcondition}
  \KwIn{$\mathcal{O}$, $T$}
  \KwIn{$MO$, $C$, $k_{max}$, $lb$, $ub$, $stop$}
  \KwOut{List of mutants}

  \SetKwProg{Def}{def}{:}{}
  \Def{BinarySearch($lb$, $ub$)}{
    $m \leftarrow (lb + ub) / 2$\\
    $SetParam(C, m)$\\
    
    $k \leftarrow 1$\\
    $L \leftarrow \varnothing$\\
    \While{$k \le k_{max}$}{
      $mut \leftarrow GenerateMutant(\mathcal{O}, MO, C, k)$\\
      $L \leftarrow L \cup Evaluate(mut, T)$\\
      $RSE \leftarrow CalculateStats(L)$\\

      \If{$IsStable(RSE)$}{
        \BREAK\\
      }
      $k \leftarrow k + 1$\\
    }
    
    \uIf{Failed to find stable $k$}{
      $ub \leftarrow m$\\ 
    }\uElseIf{$IsKilled(mut)$}{
      \tcp{Archive the mutant if it is killed.}
      $Archive(mut)$\\
      $ub \leftarrow m$\\ 
    }\Else{
      \tcp{Search higher range if not killed.}
      $lb \leftarrow m$\\
    }
    
    \uIf{$stop$ criteria are met}{
      \Return{}\\
    }\Else{
      \Return{$BinarySearch(lb, ub)$}\\
    }
  }
\end{algorithm}

\subsection{Stable and Sensitive Mutant Generation with Binary Search}
Algorithm~\ref{algo:bs} presents a binary search algorithm designed to find killable, but non-trivial, mutants for a given MO. The binary search looks for a parameter value of the given MO that makes the mutant killable, while being at the same time very close to another parameter value making the mutant non-killable, hence resulting in a challenging and likely highly sensitive mutant. The algorithm automatically computes the minimum number of instances needed to generate stable mutants. It takes several inputs, including a set of original DL models $\mathcal{O}$, a dataset $T$, a mutation operator $MO$, and the MO parameter $C$ to be searched by binary search (e.g., a ratio). The remaining inputs are the bounds of the search space of the parameter values, $lb$ and $ub$, along with the maximum number of instances, $k_{max}$, considered when searching for a stable mutant. The last input is the $stop$ criterion (e.g., timeout), which decides the budget of the search.

Initially, the binary search assigns the middle value of the current bounds to the chosen MO parameter (Lines 2-3). Next, in each iteration (Line 6), it applies the MO to generate a set of mutant instances from the original models in $\mathcal{O}$ (Line 7), producing a total of $|\mathcal{O}|$ new mutant instances per iteration. The algorithm then evaluates these new instances and checks the RSE of all generated instances so far. If the RSE is less than the user-defined threshold (checked by $IsStable$ function at Line 10), the iteration stops. Otherwise, it continues generating additional instances up to the maximum of $k_{max}$ iterations, meaning that the maximum number of mutant instances can be $k_{max} \times |\mathcal{O}|$.
Failure to find a stable mutant suggests the need for adjusting the search to a narrower range to identify a less impactful parameter value (Lines 13-14). In fact, we operate under the assumption that higher parameter values produce a higher impact (change) on the model. This is true of most parameters of \dm and of the ratio of our new operators. With a high MO parameter value, the mutant's behaviour becomes less predictable across multiple applications of the MO, as the affected parts of the model may be more or less critical for the predictions. On the contrary, lower values make the mutant increasingly more similar to the original, hence replicating the stability of the latter.
In cases where a stable mutant is found, the algorithm checks for its killability by the input dataset $T$ (see Equation~\ref{eq:kill}). If killed, the mutant is archived and the search range is narrowed into smaller value ranges to find a mutant that can deviate minimally from the original model, hopefully resulting in a hard-to-kill mutant (Lines 15-17). If the mutant survives, the search range is adjusted to larger value ranges to produce a larger deviation from the original model (Lines 18-19). The search is terminated when the stopping criteria are met (Lines 20-21). At the end of the binary search, we are left with a subset of mutants that are both stable and killable w.r.t. the given input dataset $T$, ranked from easier (found initially) to more challenging (found at the end of the search) to kill.

While any input dataset $T$ can be used for guiding the search and killability assessment, to obtain an accurate estimation of this property we need a dataset that the model can process with high accuracy, because such a dataset would expose even minor corruptions of the model. For this reason, we prioritise the original training set for the initial binary search in our experiments. The output of this search forms the basis for the killability assessment. This aligns with how we calculate the mutation score and sensitivity, both of which depend on the training set's killability (see Equations~\ref{eq:ms} \& \ref{eq:sens}). %
We apply the binary search to our new operators, as well as to the MOs of \dm, ensuring a fair comparison among tools, although this means that we are using an improved version of \dm. This is because the original tool does not ensure killability or stability, neither does it search for challenging mutants.

\subsection{\new{Inhibitor Mutation Operators}}

We suspect the low sensitivity of existing post-training mutation tools~\cite{NHGJPT21} to be due to their aggressiveness and limited control over the degree of mutation. For example, NEB blocks the weights in a binary manner and NAI flips the sign of the weights, which we believe can cause unexpected drastic changes to the model. Such mutants can be killed by any test set and therefore they are not useful for differentiating between test sets of varying quality.
While GF allows for a more nuanced adjustment of weights by applying a variable amount of Gaussian noise, it can also flip the sign of the weights when the multiplicative noise factor becomes negative.
Specifically, GF is implemented by \dm as shown in the following Equation~\ref{eq:gf}:

\begin{equation}
    \label{eq:gf}
    \begin{aligned}
    GF(W, \rho, \sigma) &= \text{Clip}\left( w \cdot (1 + \epsilon) \right) \\
    \epsilon \sim \mathcal{N}(0, \sigma^2) ~&~~~~~ w \sim \text{Uniform}(W, \rho)
    \end{aligned}
\end{equation}

\noindent
where the weights $w$ to be mutated are sampled uniformly among the model weights $W$ with probability $\rho$.
The noise value $\epsilon$, sampled from $\mathcal{N}(0, \sigma^2)$, ensures that weights are changed by a small percentage. However, when $\epsilon$ is lower than $-1$, the result is a weight sign flip, which is very impactful. 
Another problematic aspect of \dm's implementation of GF is the clipping of the mutated weights within the range of $-1$ to $1$. We speculate that this decision was aimed at preventing extremely large or small weights. On the other hand, it might restrict too much the change rate, resulting in non-killable, likely equivalent mutants that are detrimental to the sensitivity of the tool to the test set quality.

\new{To address these concerns, we propose two novel post-training mutation operators named Weight Inhibitor (WI) and Neuron Inhibitor (NI). The former reduces randomly selected weights across all neurons by a given factor, hence reducing (``inhibiting'') proportionally the propagation of the signal between pairs of neurons. The latter targets specific neurons and alters only their outgoing weights. Both WI and NI operate with two parameters: a ratio $\rho$ that controls the number of weights to be affected, and an inhibition factor $\lambda$ that controls the degree of weight mutation.}
WI can be formally described as in Equation~\ref{eq:wi}:

\begin{equation}
    \label{eq:wi}
    \begin{aligned}
    WI(W, \rho, \lambda) &= w \cdot (1 - \lambda) \\
    w & \sim \text{Uniform}(W, \rho)
    \end{aligned}
\end{equation}

\noindent Given the ratio $\rho$ and inhibition factor $\lambda$, where both are bounded within the range [0-1], WI first randomly selects $(100 \times \rho)\%$ of the weights, denoted as $w \sim \text{Uniform}(W, \rho)$, and then mutates the selected weights with the inhibition factor: $w' = w \cdot (1 - \lambda)$. Notably, if the inhibition factor is at its maximum value of 1, the selected weights are blocked, resulting in having similar effects as NEB.

\new{Similarly, NI randomly selects a proportion of neurons (i.e., $(100 \times \rho) \%$) from the set of all neurons and operates on the weights within these selected neurons. As with WI, the chosen weights are mutated by a factor of $(1 - \lambda)$.}

As inhibitor operators always reduce the weights without altering their signs, they prevent the weights from flipping and from becoming too large, the latter being the reason why \dm introduced clipping in the GF operator. By supporting the binary search for challenging $\lambda$ values, inhibitor operators offer better weight control compared to existing post-training MOs, enhancing the sensitivity of the mutants.

\section{Empirical Study with \name}
\label{sec:study2}
We conduct a comparative analysis of \name, \dm, and \dc, focusing on their sensitivity to test data quality variations and their overall stability during the generation of mutants. Moreover, we present the spectral analysis of the similarities between mutants generated with \name and those of \dc.

\subsection{Research Questions}

\subsubsection{\textbf{RQ3 (Comparison)}} \textit{How does \name perform in terms of generating sensitive and stable mutants compared to \dm and \dc?}

We compare the sensitivity and stability of the mutation testing tools. For sensitivity evaluation, we use two test sets for each subject, a strong and a weak one. Both are taken from the \dc artifact\footnote{\url{https://zenodo.org/record/4772465}}: the strong set is identical to the original test set, %
while the weak set is created by removing inputs with low confidence in the case of classification problems and by removing inputs with low mean loss/standard deviation of loss for regression problems, so that low-confidence, corner-case inputs, possibly useful to kill a mutant, are removed.

For stability comparison, we measure the number of instances needed to achieve an RSE value below a pre-defined threshold (5\%). We report stability based on all candidate mutants explored during the search, including those that would be eliminated due to non-killability or instability, because we want to measure the intrinsic stability of tools before any filters are applied. A difference from RQ2 is that we do not have a fixed parameter value for a given MO, since such values are now dynamically chosen during the search.

\subsubsection{\textbf{RQ4 (Spectral Similarity)}} \textit{How similar are the mutants generated by \name compared to those of \dc based on spectral analysis?}

We investigate whether the mutants generated with \name are diverse w.r.t. the ones of \dc. For this, we adopt \textit{spectral analysis}~\cite{spectra2024}, which estimates the similarity between two given trained neural networks (or between two sets of model instances) %
by calculating the distance between spectra of these models. The \textit{spectrum} of a model is defined as the probability distribution of the activation values of its neurons. Humbatova et al.~\cite{spectra2024} showed that such a representation of a model's behaviour is more reliable than simply referring to raw activation values since the activation values are prone to change across different trainings of the same model.

We first compare the spectra within each tool (i.e., we compute distances between pairs of mutants of either \name or \dc,  originated from the same MO but with different parameter values) and then between \name and \dc (i.e., distances between a \name's mutant and a \dc's mutant). By comparing the distribution of distances within and between tools, we can draw conclusions about (1) the diversity of mutants produced by each tool; and (2) the diversity of one tool's mutants compared to the other's. 

Following the methodology proposed by Humbatova et al.~\cite{spectra2024}, we sampled 10\% of test data for each subject. %
For each input and mutant pair, we average the spectra calculated across all the available mutant instances. We then calculate the distances between pairs of mutants based on their input-based spectra. The distances across all inputs are averaged to produce the final similarity measure.

\subsection{Evaluation Metrics}
\label{sec:eval_metrics}

\subsubsection{Sensitivity}
\label{sec:sensitivity}
\new{We use the sensitivity as a metric to evaluate the quality of the mutants, as introduced in Section~\ref{sec:pre-training}.} Note that sensitivity calculation is not possible when no killable and stable mutants are generated under the guidance of the training set. Conversely, if mutants are successfully generated using the training and strong sets but not the weak set, a perfect sensitivity (i.e., 1.0) is assumed. This is because this case clearly shows the weak set's ineffectiveness in guiding the generation of killable, non-trivial mutants.

\subsubsection{Spectral Distance}
We use Log-Euclidean metric to compute spectral distances for a pair of mutants, as it helps to avoid the cases where extremely high spectral values prevent the influence of smaller values on the average distance calculation.

\subsection{Implementations and Configurations}

We built \name on Python 3.8.10 and Keras with Tensorflow 2.8.0. %
The experiment configuration and datasets remain identical to those described in Section~\ref{sec:study1}. If there are multiple MO parameters, %
we select one parameter for the search while keeping the other ones set to their default values. Mutant generation with binary search utilises 20 original trained models. $k_{max}$ is set to 5, the RSE threshold is set to 0.05, and the search stops when the range size shrinks to a predefined precision degree of $5 \times 10^{-4}$. %

\begin{figure}[h]
  \centering
  \begin{subfigure}{0.5\linewidth}
      \centering
      \includegraphics[width=\linewidth]{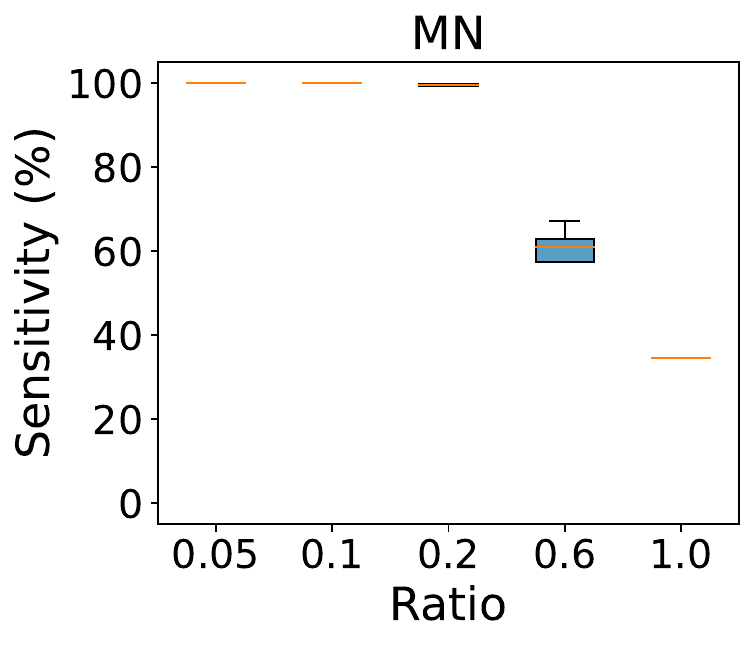}
      \label{fig:mnist_inhibitor_sens}
  \end{subfigure}%
  \begin{subfigure}{0.5\linewidth}
      \centering
      \includegraphics[width=\linewidth]{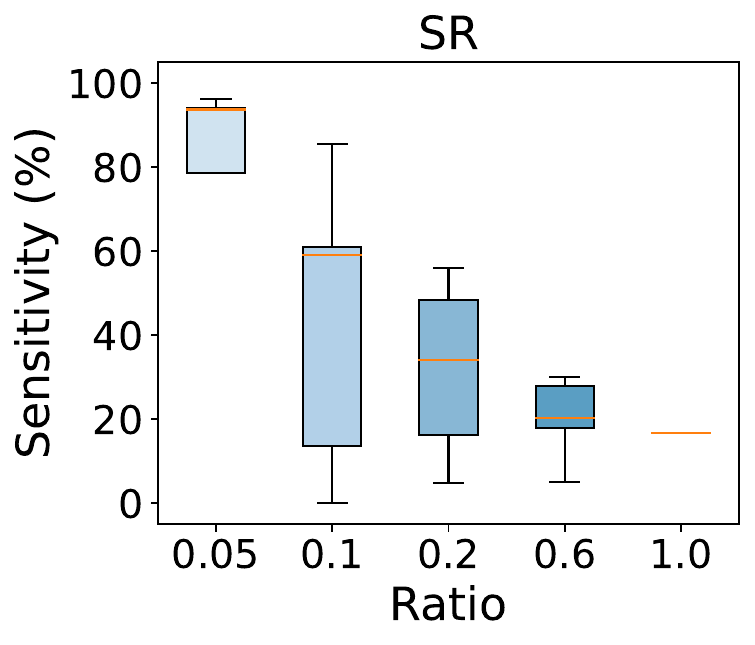}
      \label{fig:sr_inhibitor_sens}
  \end{subfigure}
  \begin{subfigure}{0.5\linewidth}
      \centering
      \includegraphics[width=\linewidth]{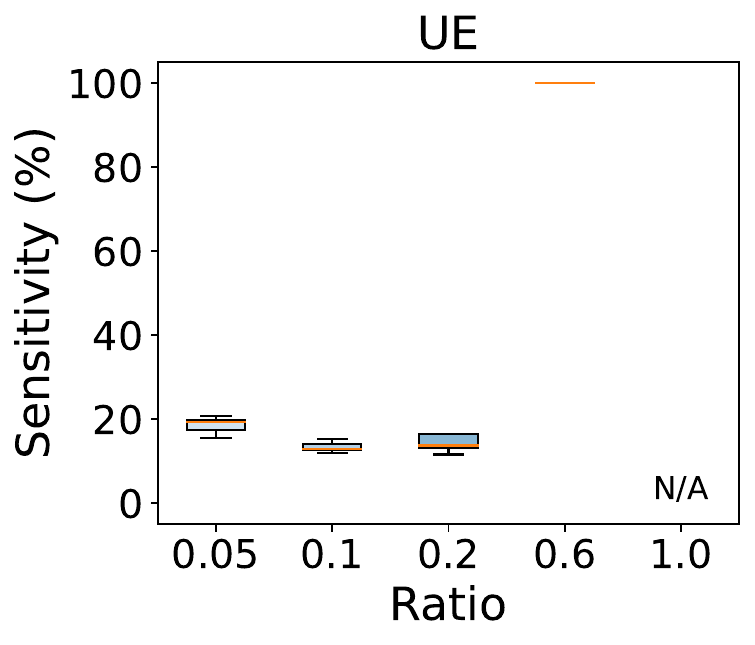}
      \label{fig:ue_inhibitor_sens}
  \end{subfigure}%
  \begin{subfigure}{0.5\linewidth}
      \centering
      \includegraphics[width=\linewidth]{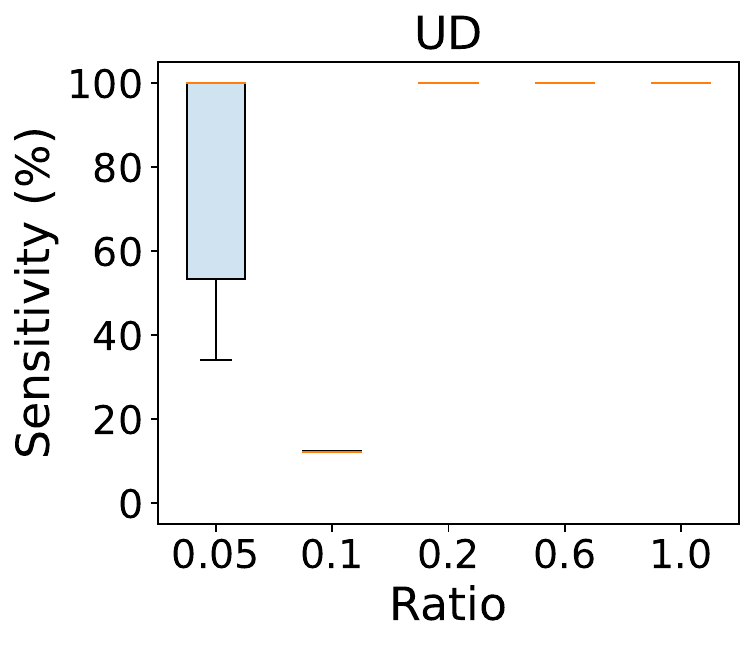}
      \label{fig:ud_inhibitor_sens}
  \end{subfigure}
  \caption{Impact of varying parameter values of WI's \textit{ratio} on the sensitivity of the generated mutants.}
  \label{fig:inhibitor_sensitivity}
\end{figure}

We ran some preliminary experiments to configure the parameter $\rho$ (ratio) of WI to a fixed value, as only $\lambda$ is subjected to binary search in our study. 
Figure~\ref{fig:inhibitor_sensitivity} shows the results of these preliminary experiments. 
We tested ratios of 0.05, 0.1, 0.2, 0.6, 1.0. Smaller ratios generally lead to higher sensitivity, likely because stronger mutations, obtained at larger ratios, tend to generate mutants that are trivial to kill, hindering the subtle changes that are necessary to achieve high sensitivity. Thus, 0.05 is chosen as the default ratio for further experiments. 
Notably, perfect sensitivity is achieved in some cases. These cases represent scenarios where the weak test set does not guide mutant generation due to two reasons: (1) for MN, all candidate mutants are stable but not killed by the weak test set and (2) for UE and UD, none of the candidate mutants are stable. Although the latter case makes an exception to the overall trend, we can attribute this to the impact of larger ratios: these ratios may affect a wider range of weights, potentially hindering the generation of stable mutants for specific models.

\subsection{Results}

\subsubsection{RQ3 (Comparison)}
\label{sec:RQ3}

\begin{table}[h]

\centering
\caption{Sensitivity of mutants: underline indicates highest sensitivity across MOs; boldface highest average sensitivity across tools.}\label{tab:sensitivity}

\centering
\begin{tabular}{c|c|c|c|c|c}
\toprule

Tool & Operator & \multicolumn{4}{c}{Subject} \\
     &          & MN & SR & UE & UD \\
\midrule
\multirow{9}{*}{\rotatebox[origin=c]{90}{\dm}} & GF & 96.78\% & 25.91\% & 1.27\% & 40.09\% \\
 & WS & 0.29\% & 2.14\% & 0.57\% & U/NK \\
 & NEB & 1.15\% & 2.68\% & 0.97\% & U/NK \\
 & NAI & 1.14\% & 1.44\% & U/NK & U/NK \\
 & NS & 0.01\% & 1.89\% & 0.67\% & 40.00\% \\
 & LR & U/NK & N/A & N/A & U/NK \\
 & LA & U/NK & N/A & N/A & U/NK \\
 & LD & U/NK & U/NK & N/A & U/NK \\\cmidrule{2-6}
 & Avg. & 19.87\% & 6.81\% & 0.87\% & 40.04\% \\
 \midrule

\multirow{25}{*}{\rotatebox[origin=c]{90}{\dc}} & TCL & 86.00\% & 0.00\% & 14.00\% & 0.00\% \\
 & TRD & 92.00\% & 66.00\% & U/NK & U/NK \\
 & TUD & 90.00\% & N/A & U/NK & N/A \\
 & TAN & \underline{100.00\%} & N/A & U/NK & N/A \\
 & TCO & 0.00\% & 3.00\% & N/A & 0.00\% \\
 & HBS & \underline{100.00\%} & N/A & U/NK & N/A \\
 & HLR & \underline{100.00\%} & N/A & \underline{51.00\%} & \underline{100.00\%} \\
 & HNE & \underline{100.00\%} & 60.00\% & U/NK & \underline{100.00\%} \\
 & HDB & \underline{100.00\%} & N/A & N/A & N/A \\
 & ACH & \underline{100.00\%} & \underline{100.00\%} & U/NK & N/A \\
 & ARM & U/NK & N/A & U/NK & N/A \\
 & AAL & N/A & N/A & 14.00\% & N/A \\
 & RAW & \underline{100.00\%} & N/A & U/NK & N/A \\
 & RCW & N/A & N/A & N/A & N/A \\
 & RRW & N/A & N/A & N/A & N/A \\
 & RCD & N/A & N/A & N/A & N/A \\
 & RCP & N/A & 43.00\% & N/A & N/A \\
 & WCI & 50.00\% & \underline{100.00\%} & U/NK & N/A \\
 & WAB & N/A & N/A & N/A & N/A \\
 & WRB & N/A & 0.00\% & N/A & N/A \\
 & LCH & \underline{100.00\%} & U/NK & U/NK & U/NK \\
 & OCH & 67.00\% & 50.00\% & U/NK & U/NK \\
 & OCG & N/A & N/A & N/A & N/A \\
 & VRM & N/A & N/A & N/A & N/A \\\cmidrule{2-6}
 & Avg. & 84.64\% & 46.89\% & 26.33\% & 50.00\% \\

\midrule
\multirow{3}{*}{\rotatebox[origin=c]{90}{\name}} & WI & \underline{100.00\%} & 83.04\% & 18.56\% & 77.50\% \\

& NI & \underline{100.00\%} & \underline{100.00\%} & 
39.28\% & \underline{100.00\%} \\\cmidrule{2-6}

& Avg. & \textbf{100.00\%} & \textbf{91.52\%} & \textbf{28.92\%} & \textbf{88.75\%} \\

\bottomrule

 \end{tabular}
\end{table}

\begin{table*}[h]
\centering
\caption{Stability of mutants of each tool}\label{tab:stability}
\resizebox{\textwidth}{!}{%
\begin{tabular}{c|c|c|c|c|c|c|c|c|c|c|c}
\toprule

Subj. & \multicolumn{2}{c|}{\name} & \multicolumn{8}{c|}{\dm} & \dc \\ 
& WI & NI & GF & WS & NEB & NAI & NS & LR & LA & LD & Unstable MOs\\
\midrule

MN & 20 & 20 & 20 & 28 & 21 & 46 & 22 & 20 & 56 & 20 & ARM (25\%) \\
SR & 20 & 20 & 20 & 25 & 23 & 27 & 25 & N/A & N/A & 20 & LCH (92\%) \\
UE & 22 & 23 & 26 & 77 & 70 & 96 & 78 & N/A & N/A & N/A & \makecell{ARM (100\%), VRM (100\%), HBS (75\%), OCH (67\%), HNE (40\%), LCH (36\%),\\ WCI (33\%), TUD (33\%), WRB (25\%), ACH (22\%), TAN (17\%), TRD (17\%), RAW (8\%)} \\
UD & 21 & 29 & 84 & 98 & 98 & 105 & 93 & 20 & 120 & 20 & OCH (33\%), TRD (17\%), LCH (8\%) \\

\bottomrule

\end{tabular}
}
\end{table*}

The results of sensitivity analysis between tools is shown in Table~\ref{tab:sensitivity}. `N/A' denotes the cases where the MO cannot be applied to the DL model, while `U/NK' indicates that all generated mutants are unstable or not killable by the train set.

\new{Results show that two \name's MOs consistently outperform \dm's MOs across all subjects, exhibiting 60\%pt higher sensitivity on average. Notably, except for GF, most \dm MOs exhibit low sensitivity. This supports our hypothesis that finer-grained control over weight adjustments leads to effective guidance in generating sensitive post-training mutants. Compared to \dc's MOs, WI shows on average 25\%pt higher sensitivity. While there are MOs of \dc that sometimes outperform WI or NI for some subjects, they do not perform consistently across all subjects, and are inapplicable to some DL models. 
When comparing our two operators,  results indicate that NI outperforms WI. This suggests that the selective mutation of specific neurons is indeed a more effective and nuanced approach. This performance advantage may be attributed to the fact that this method is less disruptive than mutating weights across all neurons.
Overall, \name offers high sensitivity, even compared to individual MOs of \dc, but at a much lower computational cost.}

Results of stability analysis for each tool are presented in Table~\ref{tab:stability}. Values in the table indicate the average number of instances needed to produce stable mutants based on RSE. Since the original model comprises 20 instances, the minimum value is 20. It subsequently increases in increments of 20 up to 100 instances. If it fails to generate a stable mutant even with 100 instances, we assign a value of 120 for the sake of averaging on five runs, although in practice such mutants are discarded during the binary search. For \dc mutants, we only calculate the RSE of the initial 20 instances and list the unstable MOs along with the percentage of unstable configurations in parentheses, since increasing instances for pre-training mutants is by design extremely inefficient, as it would require too many trainings from scratch.

\new{Compared to \dm MOs, \name's MOs generally generate stable mutants with fewer instances. This is particularly evident for the subject UD, where WI and NI require on average  21 and 29 instances respectively, while \dm MOs require significantly many more.} Note that there are exceptions: LR and LD generate extremely stable mutants but they are also highly prone to be non-killable and are discarded during the search.
The instability of \dc's MOs is subject-dependent, with notable instability observed in subject UE. It highlights the importance of stability checks even for pre-training mutants and underscores the higher stability of \name.

\begin{tcolorbox}[boxrule=0pt,frame hidden,sharp corners,enhanced,borderline north={1pt}{0pt}{black},borderline south={1pt}{0pt}{black},boxsep=2pt,left=2pt,right=2pt,top=2.5pt,bottom=2pt]
\textbf{Answer to RQ3}:
In general, \name outperforms existing mutation tools for DL systems in generating sensitive and stable mutants. %
\end{tcolorbox}

\begin{figure}[!h]
  \centering
  
  \begin{subfigure}{0.5\linewidth}
      \centering
      \includegraphics[width=\linewidth]{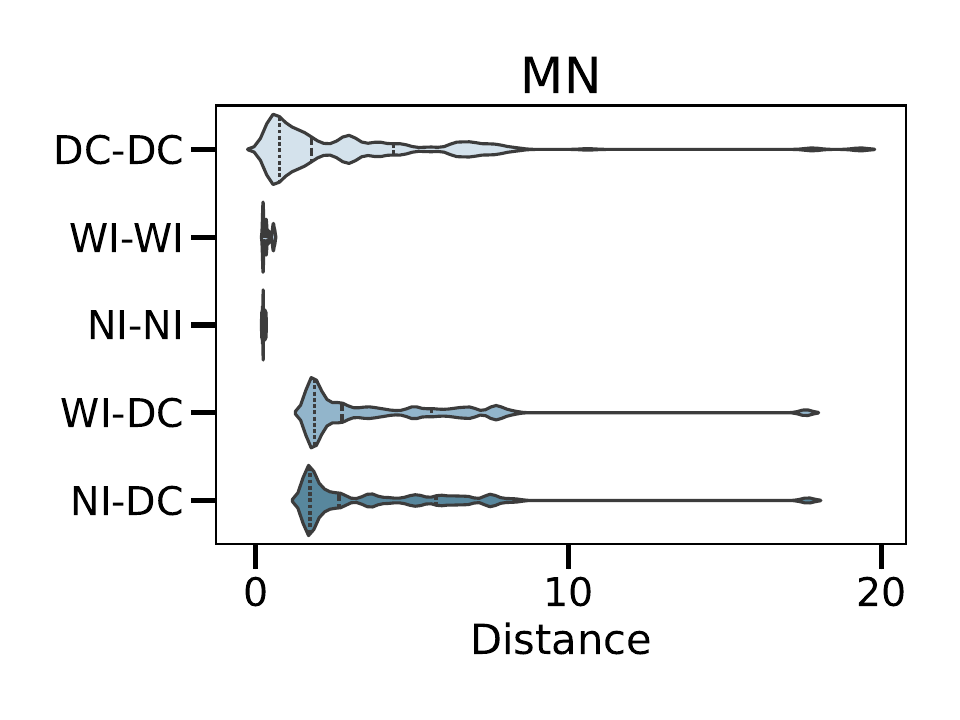}
      \label{fig:mnist_spectrum}
  \end{subfigure}%
  \begin{subfigure}{0.5\linewidth}
      \centering
      \includegraphics[width=\linewidth]{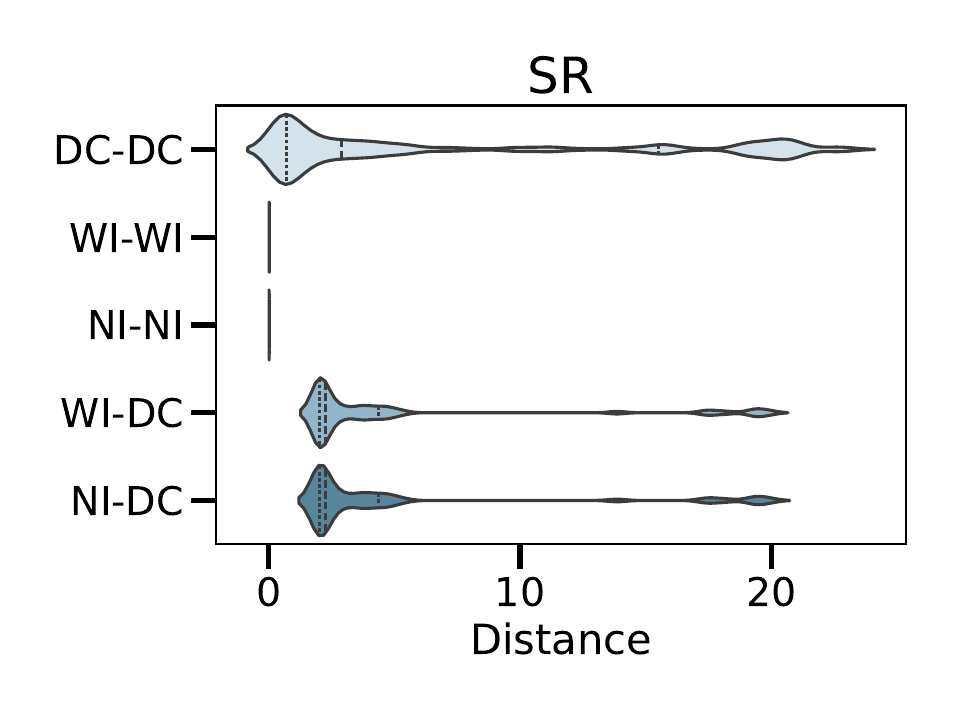}
      \label{fig:sr_inhibitor_spectrum}
  \end{subfigure}
  \begin{subfigure}{0.5\linewidth}
      \centering
      \includegraphics[width=\linewidth]{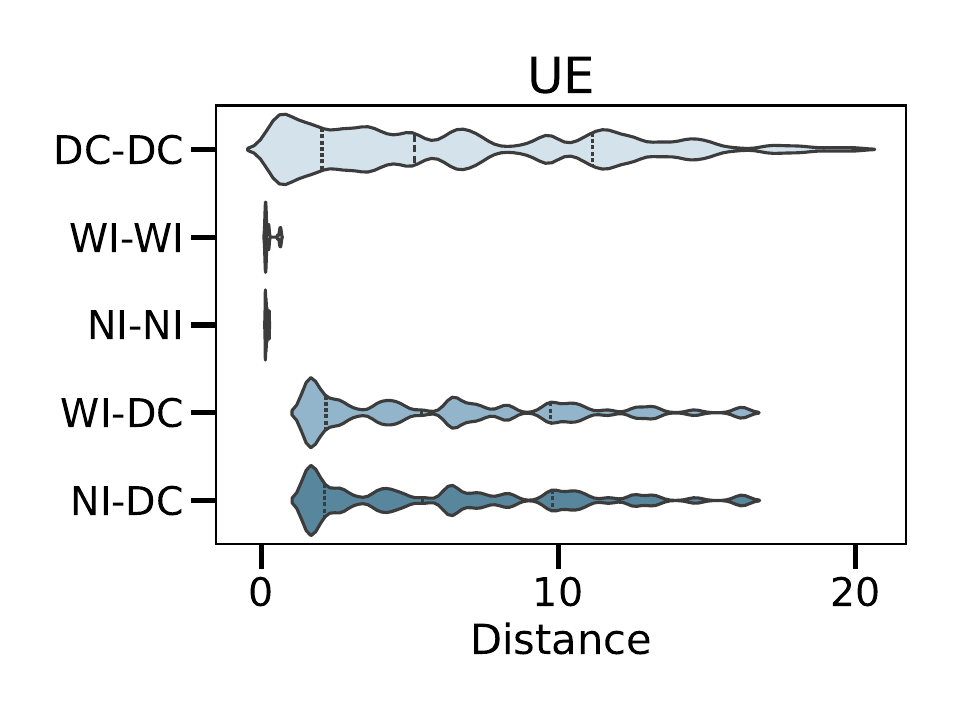}
      \label{fig:ue_inhibitor_spectrum}
  \end{subfigure}%
  \begin{subfigure}{0.5\linewidth}
      \centering
      \includegraphics[width=\linewidth]{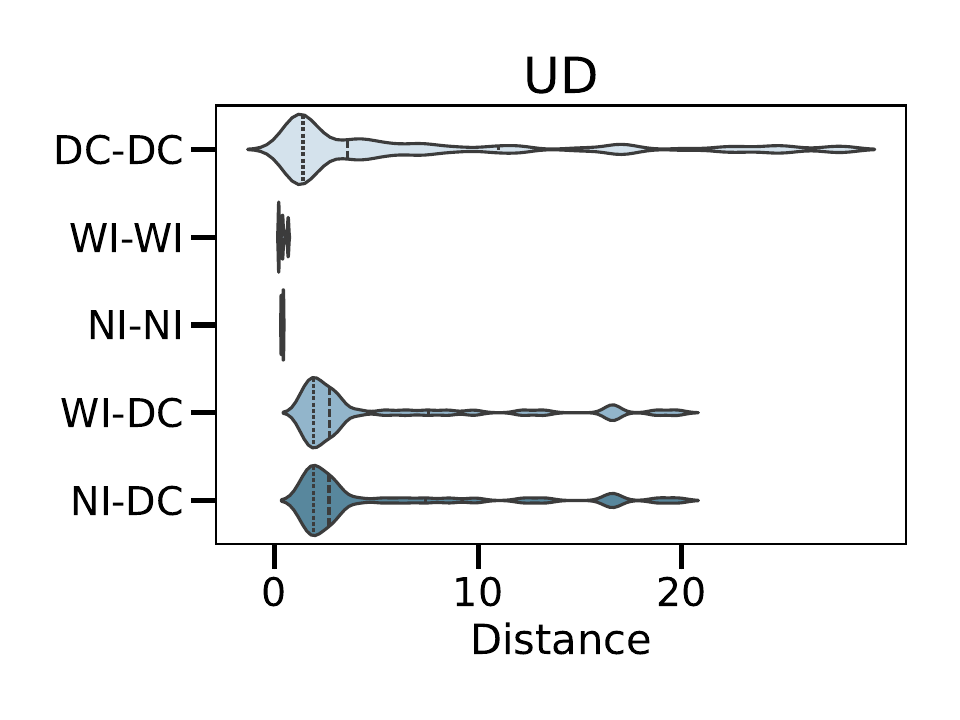}
      \label{fig:ud_inhibitor_spectrum}
  \end{subfigure}
  \caption{\new{Log-Euclidean distance distribution based on the spectrum analysis.}}
  \label{fig:rq4}
\end{figure}

\subsubsection{RQ4 (Spectral Similarity)}
\label{sec:RQ4}

Figure~\ref{fig:rq4} depicts the distribution of spectral distances using a violin plot. It shows that distances between \name mutants with WI and NI (marked as `WI-WI', `NI-NI') are \textit{always} smaller compared to distances between the \name and \dc mutants (`WI-DC' and `NI-DC'). This suggests (1) a high degree of internal consistency among \name mutants, indicating that they share characteristics, and (2) a high distinctness from \dc mutants, meaning that they represent a unique class of mutants compared to those generated by \dc. Notably, in contrast to the distances between \name mutants, the distances between \dc mutants are larger. This disparity can be attributed to \dc's MOs with categorical parameters, which can significantly differ from one another despite being defined under the same MO. For instance, the top 95\% of the largest distances between \dc mutants all originate from MOs with categorical parameters, such as LCH or ACH.

\begin{tcolorbox}[boxrule=0pt,frame hidden,sharp corners,enhanced,borderline north={1pt}{0pt}{black},borderline south={1pt}{0pt}{black},boxsep=2pt,left=2pt,right=2pt,top=2.5pt,bottom=2pt]
\textbf{Answer to RQ4}: Spectrum analysis shows that  \name mutants exhibit high internal consistency, indicating shared characteristics and are well distinguished from \dc mutants. %
\end{tcolorbox}

\section{Discussion}
\label{sec:discussion}
We further discuss the efficiency of \name compared to other tools and the variability of WI and NI.

\subsection{Efficiency}

\begin{table}[h]
\centering
\caption{Time in seconds required to generate one mutant}\label{tab:efficiency}
\scalebox{1.0}{
\begin{tabular}{c|c|c|c}
\toprule

Subj. & \name & \dm & \dc\\ 
\midrule

MN & 204 & 13 ($\downarrow$15.12x) & 1,212 ($\uparrow$6x) \\
SR & 255 & 19 ($\downarrow$13.62x) & 18,549 ($\uparrow$73x) \\
UE & 170 & 14 ($\downarrow$12.55x) & 22,445 ($\uparrow$132x) \\
UD & 726 & 44 ($\downarrow$16.38x) & 39,976 ($\uparrow$55x) \\
\midrule
Avg. & 339 & 23 ($\downarrow$15.04x) & 20,545 ($\uparrow$61x) \\

\bottomrule

\end{tabular}
}
\end{table}

While the pre-training mutation tool, \dc, has the advantage of generating mutants based on real faults, its computational cost due to training the mutants can be prohibitive. \name achieves comparable (on average even higher) effectiveness and at a much increased efficiency. 
Therefore, we investigated quantitatively the efficiency improvement achieved by \name compared to \dc and \dm. 
Note that a direct comparison to \dm can be misleading, as the latter has no control for stability, killability and triviality, which makes it inherently faster, but with lower stability and sensitivity. %

We measured the time in seconds required to generate a mutant using each tool's MOs, as shown in Table~\ref{tab:efficiency}. Values in parentheses show how much faster ($\uparrow$) or slower ($\downarrow$) \name is compared to the other tool. We observe that \name is, on average, 61 times faster than \dc but 15 times slower than \dm. While \name exhibits some slowdown compared to \dm, this trade-off ensures higher stability and sensitivity. On the other hand, the significant efficiency gain over \dc makes \name a practical and scalable solution for real-world applications.

\subsection{\new{Variability of \name's Operators}}

\begin{table}[!h]
\centering
\caption{Disagreement rate of WI, NI, and GF}\label{tab:inhibitor_variability}
\scalebox{1.0}{
\begin{tabular}{c|c|c|c|c|c|c}
\toprule

Subj. & \multicolumn{3}{c|}{1-original} & \multicolumn{3}{c}{20-originals}  \\  
& WI & NI & GF & WI & NI & GF  \\  
 
\midrule

MN & 0.43\% & 0.47\% & \textbf{0.28\%} & 1.30\% & 1.23\% & \textbf{1.11\%} \\
SR & 8.28\% & \textbf{8.22\%} & 8.63\% & 16.29\% & 16.11\% & \textbf{12.81\%} \\
UE & \textbf{19.70\%} & 65.99\% & 70.52\% & 45.76\% & \textbf{43.33\%} & 80.72\% \\
UD & \textbf{26.16\%} & 44.64\% & 33.24\%  & 27.79\% & \textbf{26.57}\% & 31.31\% \\
\midrule
Avg. & \textbf{13.64\%} & 29.84\% & 28.17\% & 22.79\% & \textbf{21.81\%} & 31.49\% \\
\bottomrule

\end{tabular}
}
\end{table}

RQ1 investigated the variability of MOs of \dm, which showed significant inconsistencies in its prediction outcomes (see Tables~\ref{tab:rq1_1} \& \ref{tab:rq1_2}). Also, we studied the stability of \name's MOs and compared it to GF, which was shown to be the most stable MO among \dm's (see Table~\ref{tab:stability}). To further investigate the variability of \name's MOs, we present the disagreement rates of WI and NI compared to GF, as shown in Table~\ref{tab:inhibitor_variability}, in which the lowest disagreement rates are highlighted in bold. While GF exhibits a slightly lower disagreement rate for MN,  WI and NI show a significantly lower disagreement rate for UE and UD. On average, in both 1-original and 20-originals scenarios,  WI or NI exhibit lower variability.

\section{Threats to Validity}
\label{sec:threats}

Threats to \textbf{internal validity} lie in the correctness of our implementation, the stochastic nature of model training, and the selection of MOs and their configurations. To alleviate them, we employed widely used DL frameworks and made our implementation available online. Furthermore, we generated all mutants considering their stability and repeated the experiment multiple times. %
Threats to \textbf{external validity} include the choice of datasets and DL models. To manage these threats, we chose widely used datasets and models with public implementation.
Threats to \textbf{construct validity} arise from the calculations of the mutation score, sensitivity, and evaluation metrics of the models. We adopted existing definitions~\cite{jahangirovantonella} and evaluation metrics that are commonly used and standard in the field.

\section{Related Work}
\label{sec:related_work}

Recently, a variety of techniques have been proposed for mutation testing in DL systems. MuNN~\cite{munn} is an early post-training mutation tool proposing five DL mutation operators such as deleting neurons or changing weight values. %
Subsequent studies include \dmm~\cite{deepmut} and its successor \dm~\cite{deepmut++}. \dmm presents source-level operators such as duplicating training data or removing activation function, as well as model-level operators (i.e., post-training operators). \dm, on the other hand, focuses on the model-level operators, expanding them with nine new operators specifically tailored for Recurrent Neural Networks (RNNs). Such operators were shown to facilitate the quantitative analysis of test data quality. \dc~\cite{NHGJPT21} is a pre-training DL mutation tool with 35 operators derived from real DL faults~\cite{taxonomy}. It employs a statistical definition of mutation killing~\cite{jahangirovantonella}. The empirical study on \dc shows that it could generate mutants that are more sensitive to the quality of the test set than the mutants of \dm.
Based on an in-depth analysis of \dm's MOs, \name advances \dm with new operators (WI and NI), a new procedure to ensure stability, and a binary search to find challenging mutants.

DeepMetis~\cite{riccio2021deepmetis} adopted \dc mutants for guiding test input generation. It aimed to increase the mutation score of a given test set by adding new inputs generated using a search-based strategy. Interestingly, post-training mutations have been used for the opposite purpose, i.e., to fix the DL model. CARE~\cite{sun2022causality} identified target neurons and mutated the weights of those neurons to fix the model. Similarly, Arachne~\cite{Sohn2022cr} directly manipulated weights using a Differential Evolution (DE) to fix the misbehaviour of the model, while keeping its correct behaviours. \name's advantages extend beyond assessing test sets. Its ability to efficiently generate sensitive mutants will make it a valuable tool for guiding test generation. \name's operators could also be potentially adapted to work with the aforementioned weight-repair methods.

\section{Conclusion}
\label{sec:conclusion}

We introduce \name, a novel post-training mutation technique for DL systems that addresses the limitations of existing methods in efficiently generating stable, killable, and sensitive mutants. To generate stable and killable mutants, \name automates mutant stability assessment through iterative generation and evaluation of mutant instances and employs binary search to efficiently identify killable mutants. Moreover, to achieve higher sensitivity, it introduces two novel operators, named Weight Inhibitor and Neuron Inhibitor, which provide finer-grained weight control compared to existing techniques. Extensive empirical evaluation demonstrates \name's effectiveness: it generates mutants with 60\%pt and 25\%pt higher sensitivity compared to \dm and \dc mutants, respectively, while surpassing \dm in generating stable mutants. Furthermore, spectral analysis shows that \name produces diverse mutants compared to those of \dc, and maintains the efficiency benefits of post-training mutation tools, being 61 times faster than \dc.

\section*{Data Availability}
The implementation of \name, data, and results are publicly available at \url{https://doi.org/10.6084/m9.figshare.28225025.v1}.

\bibliographystyle{IEEEtran}
\balance
\bibliography{bib,newref}

\end{document}